\documentstyle[11pt,psfig]{article}
\begin{document}
\def\S{${\cal S}$ }

%\begin{titlepage}
%\thispagestyle{empty}
\begin{flushright}
Utrecht-THU-96/11\\
gr-qc/9602043\\
\end{flushright}
\vskip 1cm
\begin{center}
{\LARGE\bf  Focusing and the Holographic Hypothesis}\\
\vskip 1cm
{\large Steven Corley and Ted Jacobson}
\vskip .5cm
{\it Institute for Theoretical Physics, University of Utrecht\\
P.O. Box 80.006, 3508 TA Utrecht, The Netherlands}\\
and\\
{\it Department of Physics, University of Maryland\\
                          College Park, MD 20742-4111, USA}\\
             {\tt corley, jacobson@umdhep.umd.edu}
      
\end{center}
\vskip 1cm

\begin{abstract}

The ``screen mapping" introduced by Susskind to implement 't Hooft's
holographic hypothesis is studied.  For a single screen time, there are
an infinite number of images of a black hole event horizon, almost
all of which have smaller area on the screen than the horizon area.
This is consistent with the focusing equation because of the existence
of focal points.  However, the {\it boundary} of the past (or future)
of the screen obeys the area theorem, and so always gives an expanding
map to the screen, as required by the holographic hypothesis. These 
considerations are illustrated with several axisymmetric static 
black hole spacetimes.

\end{abstract}
%\end{titlepage}
\vskip 2mm

\section{Introduction}

The generalized second law of thermodynamics\cite{Bek-GSL} is the
statement that the entropy outside event horizons plus the
Bekenstein-Hawking entropy $A/4$ (in Planck units) of all event
horizons cannot decrease. The law seems to be correct, at least in
quasistationary processes\cite{GSLproofs}.  If it is true, it must be
that $A/4$ is the most entropy that could possibly be contained in a
region bounded by an area $A$\cite{Bek-GSL, tHooft-qst, tHooft-dimred}.
There has been much debate over the past 20 years about whether or not
this bound really holds, and part of the problem in proving it is that
it is not precisely clear what the statement means. Nevertheless, there
are many reasonable arguments in support of it.

If the $A/4$ bound is indeed valid, then the the number of states is
vastly overestimated by any ordinary flat space density of states in
three dimensions.  The reason is that almost all of those states would,
because of gravity, create a black hole whose entropy is fixed at
$A/4$.  One is thus led to the {\it holographic
hypothesis}\cite{tHooft-qst,tHooft-dimred,Suss,Smol-tqft}, according to
which the true degrees of freedom are enumerated on a surface enclosing
the volume of interest, at an information density less than or equal to
one bit per Planck area.
 
Susskind proposed to implement `t Hooft's holographic hypothesis by
mapping all the points of space, by light rays that impinge
perpendicularly, onto a flat two-dimensional {\it screen} \S in a
distant asymptotically flat region.  We call this mapping the {\it
screen map}.  This particular idea was partly motivated by properties
of string theory in the light cone gauge, but the mapping between
surface and volume degrees of freedom is not really  specified in
Susskind's proposal. The  proposal seems more intended to get some kind
of picture on the table so one can begin thinking about it. The first
test to which Susskind subjected the screen map was to ask whether the
horizons of any black holes that are present are necessarily mapped
onto sets of {\it larger area} on the screen. This is required by the
holographic hypothesis since the black hole has the maximal bit density
of one per unit area.  We call a screen map with this property an {\it
expanding map}.  Susskind argued that his screen map is indeed
expanding. He also noted that it is a one to many map from horizon
to screen, and suggested that at the quantum level one should perhaps 
superpose all of the images.

In this paper we take a closer look at the definition and properties of
the screen map.  We find that, for a single screen time, there are an
infinite number of multiple images, corresponding to {\it different}
horizon times. Since they arise from different time slices of the
horizon, it does not seem appropriate to superpose them. Moreover,
almost all of these single image maps are {\it contracting} rather than
expanding, which is allowed by the focusing equation on account of the
presence of focal points.  However, we prove in general and illustrate
in several examples that there is always at least one expanding image
of the horizon. This proof is essentially Hawking's area theorem,
applied to the boundary of the past (or future) of the screen.

\section{The screen map}

We are interested in light rays that hit the screen orthogonally at one
screen time, that is, at one particular spacelike slice \S of the
screen's history. The first thing to clarify is whether the light
rays leave the screen towards the future or towards the past. Lacking
the holographic theory, we do not have a way to decide this, so we
shall consider both possibilities.  If the rays from the screen are
{\it past directed}, then they never actually cross the future black
hole horizon.  They can however cross the ``stretched horizon"
\cite{Membrane, Suss-stretch}, which in any case might be argued to be
more relevant than the true event horizon. If the rays from the screen
are {\it future directed}, then they can indeed cross the future black
hole horizon,  so we get a precise map from a subset of \S to the
horizon by following the rays.

We define the {\it future screen map} to be the past directed
congruence of null rays orthogonal to \S.  A point on a given ray is
mapped to the point where the ray hits the screen.  Similarly we define
the {\it past screen map} reversing the roles of past and future. In a
static spacetime, such as the maximally extended Schwarzschild metric,
these are equivalent. In particular, although the future screen map
does not intersect the future horizon, it does intersect the past
horizon. In such cases the past horizon can serve as a surrogate for
the stretched horizon.  Of particular interest will be the part of this
congruence that lies on the {\it boundary} $\dot{I}^-[{\cal S}]$ of the
past of \S.  We refer to this part as the {\it primary} future screen
map, and similarly for the past screen map.

\begin{figure}[tb]
\centerline{
\psfig{figure=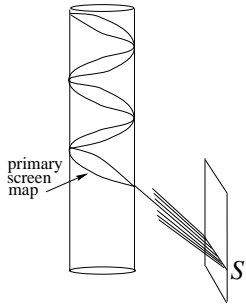,angle=-90,height=4cm}}
\caption{\small Past screen map in black hole spacetime, showing 
intersection with horizon. The multiple coverings of the horizon 
come from rays that orbit the hole before crossing the horizon.  
Only the first covering lies in the primary screen map.}
\label{screenmap}
\end{figure}

The definition of the past screen map is illustrated in Fig.
\ref{screenmap} for the case in which a single black hole is present.
The intersection of the screen map with the horizon is shown. The
multiple coverings of the horizon come from rays that orbit the hole
before crossing the horizon.  Only the first covering lies in the
primary screen map.

Rather than taking the screen to be planar, it might be natural to
think of it as a large sphere in an asymptotically flat spacetime,
which can be taken all the way out to future or past null infinity.
The distinction between planar and spherical screens makes no
difference for our general arguments (as long as the rays orthogonal
to the screen are not diverging), but the specific examples
considered below will refer to a planar screen at infinity.

\subsection{Focal points and expanding maps}

In \cite{Suss} it was argued that the screen map is necessarily expanding
as a consequence of the {\it focusing equation}
\begin{equation}
\rho'=\rho^2/2 + \sigma_{ab}\sigma^{ab} + R_{ab}k^ak^b,
\end{equation}
which governs the rate of change with respect to affine parameter of
the convergence $\rho$ of a congruence of null geodesics. $\sigma_{ab}$
is the shear tensor, $R_{ab}$ is the Ricci tensor, and $k^a$ is the
tangent vector to the congruence.  According to the Einstein equation
one has $R_{ab}k^ak^b=8\pi G T_{ab}k^ak^b$, so the rate of change
$\rho'$ is positive as long as the null energy condition
$T_{ab}k^ak^b\ge0$ holds. Thus, assuming the null energy condition, one
knows that if the light rays have vanishing convergence at the screen,
$\rho$ must be non-negative everywhere from screen to horizon
(with affine parameter increasing from screen to horizon), 
so the screen map is expanding.

There is a serious flaw in this reasoning however since $\rho$ may
become infinite somewhere, after which point it can be {\it negative}.
The focusing equation still says it must increase after that, but that
is of no help in establishing the expanding character of the map. A
point where $\rho$ is infinite is a {\it focal point} (also often
called a ``caustic" or a ``conjugate point" to the screen). To be sure
of the expanding character of the map one must show that there are no
focal points between the screen and the horizon.

\subsection{Primary screen map and the area theorem}

The {\it boundary} of the past of the screen $\dot{I}^-[{\cal S}]$ is
what we have called the primary screen map.  This boundary is similar
to a black hole event horizon, and shares with such a horizon the
property that the area of its cross sections cannot decrease toward the
future\cite{HawkElli}.  This property follows from the focusing
equation, the null energy condition, and a key property of all past
boundaries: each point lies on a null geodesic that runs all the way up
the boundary to \S with no focal points along the way.  The proof
assumes that no ``naked singularities", i.e. singularities visible from
the screen, are encountered along the way up to \S.

Thus one has an ``area theorem" for the primary screen map, which
guarantees that this map is expanding. In particular, if the null rays
cross a stretched horizon, then the stretched horizon will necessarily
be mapped to a larger area on the screen.

All the above applies equally well to the boundary of a past screen
map, with the roles of future and past interchanged. The boundary has
an area that must {\it decrease} toward the future, but still {\it
increases} toward the screen, so again one obtains an expanding map to
the screen (assuming now that no singularities are encountered on the
way from the screen).

\section{Static axisymmetric examples}

In this section we look at a number of specific examples that
illustrate the general principles already discussed.  In some of the
examples we can actually calculate the areas of the screen images of
the black hole horizon and show that a finite number (often only one)
are greater but the remaining infinite number are less than the horizon
area. We explain this fact by identifying the focal points. We also
identify the primary screen map in all the examples and show explicitly
that it has no focal points.  Since these are static spacetimes the
future and past screen maps are equivalent. For definiteness, we adopt
here the past screen map, which intersects the future black hole
horizon.

\subsection{Single black hole}

\begin{figure}[tb]
\centerline{
\psfig{figure=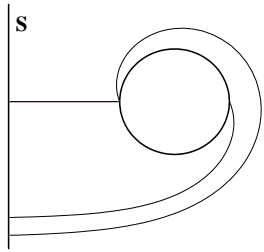,angle=-90,height=4cm}}
\caption{\small Three orbits from screen to a Schwarzschild black hole,
drawn in $(r,\phi)$ plane.}
\label{threeorbits}
\end{figure}

\begin{figure}[tb]
\centerline{
\psfig{figure=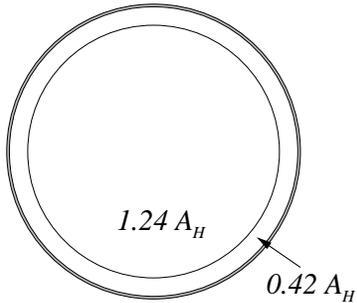,angle=-90,height=4cm}}
\caption{\small Screen image of Schwarzschild horizon. The inner disc
is the primary cover, and all the rest of the covers are annuli. The
higher order covers accumulate at an impact parameter of $3^{3/2} M$,
the capture radius. Only the first and second covers are shown explicitly. 
The rest are too narrow to show to scale. Even the second cover has
smaller area (0.42 $A_H$) than the horizon.}
\label{bullseye}
\end{figure}

For a single spherically symmetric black hole we can easily analyze the
screen map explicitly by use of the integrated geodesic equation.
Three orbits from the screen to a Schwarzschild hole are shown in Figure
\ref{threeorbits}.  The orbits can be labeled by the angle at which
they hit the horizon, measured from the perpendicular from the screen
to the hole. The orbits with angles between 0 and $\pi$ cover the
horizon once, when rotated around the axis. Those with angles between
$\pi$ and $2\pi$ give a second covering, and so on. There are an
infinite number of coverings of the horizon. There is an upper bound
for the impact parameter for capture by the hole, so the areas of the
coverings on the screen must converge to zero.  These coverings are
illustrated in Figure \ref{bullseye}, which is drawn approximately to
scale for the case of a Schwarzschild black hole. The primary cover has
area $\simeq 1.24 A_H$, greater than the horizon area, but all the rest
have lesser area. In fact the sum of all the rest has lesser area, 
$0.45 A_H$.

\begin{figure}[tb]
\centerline{
\psfig{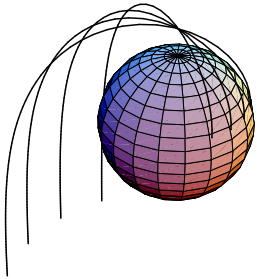}}
\caption{\small Focal point for the second cover of the screen map
to a Schwarzschild horizon.}
\label{focalpt}
\end{figure}

Since the higher order covers are not expanding, there must be at least
one focal point somewhere on each of the corresponding null congruences.
This focal point is shown in Figure \ref{focalpt} for the second cover
in the Schwarzschild case.  All the null rays will intersect at a point
directly behind the black hole (as viewed from the screen). At this
point the convergence goes to infinity and just past this point it is
negative (although in general the convergence may become positive
again and may even diverge again if the congruence goes
through another focal point).
It is interesting to note that the separation of the null
rays in the transverse direction is larger on the screen than on the
horizon, so it is not the angular focusing but the radial focusing that
makes the map fail to be expanding, even though it is the angular
focusing that produces the focal point.

We also analyzed the screen map for an extremally charged
Reissner-Nordstrom hole. There it turns out that both the primary and
the second cover are expanding, but the rest are not.

\subsection{Two distant black holes}

\begin{figure}[tb]
\centerline{
\psfig{figure=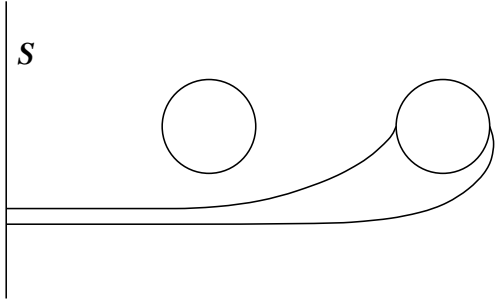,angle=-90,height=4cm}}
\caption{\small Extreme rays of the primary cover of the second black hole.
None of the rays in this cover cross the axis.}
\label{2holes}
\end{figure}

It was pointed out in \cite{Suss} that one black hole cannot be hidden
behind another, and that the screen map will still be expanding for
both horizons in this case. We investigated this situation explicitly
for the case where the two black holes are very far from one another
and the screen is perpendicular to the axis joining them.  In this case
we could approximate the capture orbits for the second hole analytically by
simply composing with the scattering by the first hole. The screen map
pattern in this case is rather complicated.  A general orbit can
alternate between the two holes, wrapping any number of times around on
each visit, before finally crossing the horizon of one of the holes.
The primary map is given by the orbits that never cross the axis (see
Figure \ref{2holes}), and it traces out an annulus on the screen. We
estimated the area of this annulus for large separation $d$, and found
that it grows like $d^{1/2}$: $A \sim (M_{1} d/M_{2}^{2})^{1/2} A_{2}$.
Here $M_{1}$, $M_{2}$ are the masses of the first and second holes
respectively,
$A_{2}$ is the horizon area of the second, and we have assumed that
$d \gg M_{2}^{2}/M_{1}$. 

\subsection{General static axisymmetric spacetime}

For two nearby black holes one cannot just compose the asymptotic
scatterings, and also one must decide what metric to take for the two
black holes.  In order to have a static situation, it is natural to
consider a pair of extremally charged black holes with like sign
charges, one of the Majumdar-Papapetrou solutions\cite{MP}.
Generalizing this, one can consider a linear array of any number of
such black holes.  We shall now show by explicit consideration of the
geodesics that the primary map has no focal points, so the map is
expanding.

It is most natural to discuss this problem for an arbitrary static
axisymmetric geometry with the screen perpendicular to the axis.
Because of the axisymmetry, focal points will occur on any rays that
cross the axis, so we restrict attention to those rays that do not
cross the axis. These can only have focal points if they are focussed
to crossing points in the radial direction.  To establish the absence
of such crossing points we adopt a technique from the analysis by
Yurtsever\cite{Yurt} of chaos in the orbits around a pair of extremal
black holes.

Consider a line element of the form $g=fdt^2-gd\phi^2 -h_{ij}dx^idx^j$,
where $i=1,2$, and $f,\, g$ and $h_{ij}$ are functions only of $x^i$.
The rays from the screen have constant $\phi$, and, since they are null
geodesics, they are the same as for the conformally rescaled
three-dimensional line element $\tilde{g}=dt^2 - \tilde{h}_{ij} dx^i
dx^j$, where $\tilde{h}_{ij}=f^{-1}h_{ij}$. The geodesics of
$\tilde{g}$  project to geodesics of the Riemannian metric
$\tilde{h}_{ij}$, and the arc length along the projection agrees with
the affine parameter along the null geodesic. Now computation reveals
that the curvature of the two-dimensional metric $\tilde{h}_{ij}$ is
negative everywhere outside the horizon of a general multi extremal
black hole solution, as well as for a Reissner-Nordstrom black hole
with any charge to mass ratio.  Thus in all these cases the projected
curves are receding from each other. The spacetime null geodesics are
therefore also receding with respect to $\tilde{g}$. While this does
not imply that they are receding with respect to the physical metric
$g$, it does at least imply that they will not reach a focal point.

\section{Remarks}

The main lesson of this geometric exercise is the conclusion that, to
guarantee the expanding nature of the holographic map from horizon to
screen, one should restrict to the {\it primary} screen map, i.e. the
boundary of the past (or future) of the screen.  This boundary
generalizes the concept of a black hole event horizon and satisfies the
area theorem.

Even the primary screen map is only guaranteed to be expanding if the
null energy condition is satisfied.  In the presence of quantum fields
this condition can be locally violated by the expectation value of the
stress-energy tensor, so one should not rely on the null energy
condition. Perhaps an {\it averaged} null energy condition would be
sufficient to establish the expanding nature of the primary screen
map.  (Several recent works\cite{anec} have studied the extent to which
averaged energy conditions hold in quantum field theory.)
Alternatively, the expanding property may be lost, which may reveal
something about the holographic principle.

Our proof of the expanding nature of the primary screen map relied 
heavily on the assumption that the rays orthogonal to the screen are 
not diverging. This is guaranteed in asymptotically flat space by taking
a flat or spherical screen at infinity. 
In a closed expanding universe it seems one must choose a 
future screen map rather than a past one in order to have any chance of 
finding an expanding map from all points of space to a single screen.
Strangely, if the universe recollapses, one must then switch to a past 
screen map.
Perhaps some insight into the holographic hypothesis can be gleaned 
from an investigation of its compatibility with closed universes.

\section{Acknowledgements}

This work was supported in part by NSF grant
PHY-94-13253, a UMd graduate fellowship, and the University of Utrecht.

\end{document}